\def\be{\begin{equation}}
\def\ee{\end{equation}}
\def\ba{\begin{array}}

\def\ea{\end{array}}

\documentclass[12pt]{article}
\usepackage{amsfonts,amssymb,amsmath,graphicx}
\begin{document}
\parskip=3pt
\parindent=18pt
\baselineskip=20pt
\setcounter{page}{1}
\centerline{\large\bf Average coherence with respect to complementary measurements}
\vspace{6ex}
\centerline{{\sf Bin Chen,$^{\star}$}
\footnote{\sf Corresponding author: chenbin5134@163.com}
~~~ {\sf Shao-Ming Fei$^{\natural,\sharp}$}
}
\vspace{4ex}
\centerline
{\it $^\star$ College of Mathematical Science, Tianjin Normal University, Tianjin 300387, China}\par
\centerline
{\it $^\natural$ School of Mathematical Sciences, Capital Normal University, Beijing 100048, China}\par
\centerline
{\it $^\sharp$ Max-Planck-Institute for Mathematics in the Sciences, 04103 Leipzig, Germany}\par
\vspace{6.5ex}
\parindent=18pt
\parskip=5pt
\begin{center}
\begin{minipage}{5in}
\vspace{3ex}
\centerline{\large Abstract}
\vspace{4ex}

We investigate the average coherence with respect to a complete set of complementary measurements.
By using a Wigner-Yanase skew information-based coherence measure introduced in [Phys. Rev. A \textbf{96}, 022130, 2017],
we evaluate the average coherence of a state with respect to any complete set of mutually unbiased measurements and general symmetric informationally complete measurements, respectively.
We also establish analytically the relations among these average coherences.
\end{minipage}
\end{center}

\newpage

\section{Introduction}

Quantum coherence, as one of the most significant quantum resources, has become a hot spot in recent years ever since Baumgratz et al. \cite{Bau} introduced the mathematical framework of quantifying the quantum coherence. Based on this framework, a variety of coherence quantifiers have been proposed, such as the $l_1$ norm of coherence, the relative entropy of coherence, the distance-based coherence, the coherence formation and the robustness of coherence \cite{Bau,Strel,Tong,Chen,Rast,Winter,Nap}.
All these measures are indubitably based on two important concepts in the framework -- incoherent states and incoherent operations.

In Ref. \cite{Luo1}, Luo et al. established a quantitative link between coherence and quantum uncertainty.
By identifying the coherence of a state (with respect to a measurement) as the quantum uncertainty of a measurement (with respect to a state),
they introduced a coherence quantifier from an alternative perspective based on quantum uncertainty described by the famous Wigner-Yanase skew information \cite{WY}.
This new measure can be mathematically expressed as
\begin{equation}
Q(\rho,\mathcal{M}):=\sum_{i}I(\rho,M_{i}),
\end{equation}
where $\mathcal{M}=\{M_{i}\}$ is a positive operator-valued measure (POVM) with $M_{i}\geq0$, $\sum_{i}M_{i}=\mathbf{1}$ with $\mathbf{1}$ denoting the identity operator,
and $I(\rho,M_{i})=-\frac{1}{2}\mathrm{Tr}[\sqrt{\rho},M_{i}]^{2}$ is the skew information of $\rho$ with respect to the measurement operator $M_{i}$.
Although there is no direct connection between this new measure and the incoherent operations or incoherent states, it is indeed a bona fide quantifier for coherence, since it satisfies many nice properties such as non-negativity, convexity, decreasing under any quantum operation, etc. \cite{Luo1}.
In this paper, we call $Q(\rho,\mathcal{M})$ the measurement-based coherence for convenience.

Recently, Luo et al. \cite{Luo2} studied the average coherence over any complete set of mutually unbiased bases (MUBs) \cite{WF,Durt},
as well as the average coherence over all orthonormal bases in terms of the measurement-based coherence measure.
They proved that these two averages are equivalent by direct evaluation.
More concretely, let $\{\mathcal{B}_{m}\}_{m=1}^{d+1}$ be a complete set of $d+1$ MUBs in a $d$-dimensional Hilbert space $\mathcal{H}_{d}$,
the average coherence of $\rho$ with respect to $\{\mathcal{B}_{m}\}_{m=1}^{d+1}$ is defined as
\begin{equation}\label{CMUB}
\mathcal{C}_{\mathrm{MUB}}(\rho):=\frac{1}{d+1}\sum_{m=1}^{d+1}Q(\rho,\mathcal{B}_{m}).
\end{equation}
The average coherence over all orthonormal bases is defined as
\begin{equation}
\mathcal{C}_{\mathcal{U}}(\rho):=\int_{\mathcal{U}}Q(\rho,U\Pi U^{\dag})dU,
\end{equation}
where $\Pi=\{|i\rangle\langle i|\}$ with $\{|i\rangle\}$ a fixed basis in $\mathcal{H}_{d}$, $U\Pi U^{\dag}=\{U|i\rangle\langle i|U^{\dag}\}$,
and the integration is taken over the set of all unitary operators acting on $\mathcal{H}_{d}$.
It has been shown that \cite{Luo2},
\begin{equation}
\mathcal{C}_{\mathrm{MUB}}(\rho)=\mathcal{C}_{\mathcal{U}}(\rho)=\frac{1}{d+1}[d-(\mathrm{Tr}\sqrt{\rho})^{2}].
\end{equation}
Another important quantity is the maximal coherence \cite{Luo1},
\begin{equation}\label{CMAX}
\mathcal{C}_{\mathrm{max}}(\rho)=\frac{1}{d}\sum_{i=1}^{d^{2}}I(\rho,H_{i})=\frac{1}{d}[d-(\mathrm{Tr}\sqrt{\rho})^{2}],
\end{equation}
where $\{H_{i}\}$ is any complete orthogonal set of observables.
It is obvious that $\mathcal{C}_{\mathcal{U}}(\rho)$ and $\mathcal{C}_{\mathrm{max}}(\rho)$ are approximately equal when $d$ is large enough.
That is to say, the coherence of a state is almost maximal with respect to all orthonormal bases for high dimensional quantum systems \cite{Luo2}.

In this paper, we study the average coherence with respect to complementary measurements.
We consider any complete set of mutually unbiased measurements (MUMs) \cite{KG1} and general symmetric informationally complete measurements (general SIC measurements) \cite{GK2}, respectively. We evaluate the average coherence of a state with respect to these special types of quantum measurements.
We find that the resulted average coherence is a constant multiple (related to the given measurements) of the maximal coherence as well as the average coherence with respect to all orthonormal bases.

\section{Average coherence with respect to MUMs}

We first recall some basic notions of mutually unbiased bases and mutually unbiased measurements.
Two orthonormal bases $\mathcal{B}_{1}=\{|b_{i}\rangle\}_{i=1}^{d}$ and $\mathcal{B}_{2}=\{|c_{j}\rangle\}_{j=1}^{d}$ of $\mathcal{H}_{d}$
are said to be mutually unbiased if
\begin{equation}
|\langle b_{i}|c_{j}\rangle|=\frac{1}{\sqrt{d}} ,~~~\forall\, i,j=1,2,\cdots,d.
\end{equation}
A set of orthonormal bases $\{\mathcal{B}_{1}, \mathcal{B}_{2},...,\mathcal{B}_{m}\}$ in $\mathcal{H}_{d}$ is called a set of MUBs if every pair of the bases in the set are mutually unbiased.
It has been shown that there are at most $d+1$ pairwise unbiased bases, which is called a complete set of MUBs \cite{WF,Durt}.
However, the existence problem of complete set of MUBs for arbitrary $d$ is still open.

In Ref. \cite{KG1}, Kalev and Gour generalize the concept of MUBs to MUMs.
Two POVM  measurements on $\mathcal{H}_{d}$,
$\mathcal{P}^{(b)}=\{P_{n}^{(b)}\}_{n=1}^{d}$, $b=1,2$, are said to be MUMs if
\begin{equation}
\begin{split}
\mathrm{Tr}(P_{n}^{(b)})&=1,\\
\mathrm{Tr}(P_{n}^{(b)}P_{n'}^{(b')})&=\frac{1}{d},~~~b\neq b',\\
\mathrm{Tr}(P_{n}^{(b)}P_{n'}^{(b)})&=\delta_{n,n'}\,\kappa+(1-\delta_{n,n'})\frac{1-\kappa}{d-1},
\end{split}
\end{equation}
where $\frac{1}{d}<\kappa\leq1$, and $\kappa=1$ if and only if all $P_{n}^{(b)}$s are rank one projectors, i.e., $\mathcal{P}^{(1)}$ and $\mathcal{P}^{(2)}$ are given by MUBs.
Unlike MUBs, there always exists a complete set of $d+1$ MUMs which can be constructed explicitly \cite{KG1}.
Let $\{F_{n,b}:n=1,2,\cdots,d-1,b=1,2,\cdots,d+1\}$ be a set of $d^{2}-1$ Hermitian, traceless operators acting on $\mathcal{H}_{d}$,
satisfying $\mathrm{Tr}(F_{n,b}F_{n',b'})=\delta_{n,n'}\delta_{b,b'}$.
Define $d(d+1)$ operators
\begin{equation}
F_{n}^{(b)}=
\begin{cases}
   F^{(b)}-(d+\sqrt{d})F_{n,b},&n=1,2,\cdots,d-1;\\[2mm]
   (1+\sqrt{d})F^{(b)},&n=d,
\end{cases}
\end{equation}
where $F^{(b)}=\sum_{n=1}^{d-1}F_{n,b}$, $b=1,2,\cdots,d+1$.
Then the following operators form a complete set of $d+1$ MUMs:
\begin{equation}
P_{n}^{(b)}=\frac{1}{d}I+tF_{n}^{(b)},
\end{equation}
as long as $t$ is properly chosen such that all $P_{n}^{(b)}$s are positive, $b=1,2,\cdots,d+1,n=1,2,\cdots,d$. The parameter $\kappa$ is given by
\begin{equation}
\kappa=\frac{1}{d}+t^{2}(1+\sqrt{d})^{2}(d-1).
\end{equation}
Moreover, any complete set of MUMs can be expressed in such form \cite{KG1}.

We now investigate the average coherence with respect to MUMs in terms of the measurement-based coherence measure.
Let $\mathcal{P}_{\mathrm{MUM}}=\{\mathcal{P}^{(b)}\}_{b=1}^{d+1}$  be a complete set of MUMs with the parameter $\kappa$. Similar to (\ref{CMUB}), we need to evaluate the following quantity,
\begin{equation}
\mathcal{C}(\rho,\mathcal{P}_{\mathrm{MUM}}):=\frac{1}{d+1}\sum_{b=1}^{d+1}Q(\rho,\mathcal{P}^{(b)}).
\end{equation}

Towards the computation of $\mathcal{C}(\rho,\mathcal{P}_{\mathrm{MUM}})$, we first consider the following quantity,
\begin{equation}
Q_{\alpha}(\rho,\mathcal{P}_{\mathrm{MUM}}):=
\sum_{b=1}^{d+1}\sum_{n=1}^{d}I_{\alpha}(\rho,P_{n}^{(b)}),
\end{equation}
where $I_{\alpha}(\rho,X)$ is the generalized skew information, usually called the Wigner-Yanase-Dyson entropy (WYD entropy) \cite{Lieb}, which is given by
\begin{equation}
\begin{split}
I_{\alpha}(\rho,X)&:=-\frac{1}{2}\mathrm{Tr}([\rho^{\alpha},X][\rho^{1-\alpha},X])\\
&=\mathrm{Tr}(X^{2}\rho)-\mathrm{Tr}(\rho^{\alpha}X\rho^{1-\alpha}X),
\end{split}
\end{equation}
where $0<\alpha<1$.
It is obvious that the WYD entropy $I_{\alpha}(\rho,X)$ reduces to the skew information when $\alpha=1/2$.
Like skew information, WYD entropy has many applications in quantum information theory, especially in characterizing the quantum uncertainties \cite{DF,Ya,Lix}.
It can be seen that
\begin{equation}\label{CQ}
\mathcal{C}(\rho,\mathcal{P}_{\mathrm{MUM}})=
\frac{1}{d+1}\sum_{b=1}^{d+1}Q_{\frac{1}{2}}(\rho,\mathcal{P}_{\mathrm{MUM}}).
\end{equation}

Next, we calculate the quantity $Q_{\alpha}(\rho,\mathcal{P}_{\mathrm{MUM}})$.
From the construction of $d+1$ MUMs given above, one gets
\begin{equation}\label{q1}
\begin{split}
Q_{\alpha}(\rho,\mathcal{P}_{\mathrm{MUM}})&=\sum_{b=1}^{d+1}\sum_{n=1}^{d}I_{\alpha}(\rho,P_{n}^{(b)})\\
&=t^{2}\sum_{b=1}^{d+1}\sum_{n=1}^{d}I_{\alpha}(\rho,F_{n}^{(b)})\\
&=t^{2}\sum_{b=1}^{d+1}\sum_{n=1}^{d}\mathrm{Tr}[(F_{n}^{(b)})^{2}\rho]
-t^{2}\sum_{b=1}^{d+1}\sum_{n=1}^{d}\mathrm{Tr}(\rho^{\alpha}F_{n}^{(b)}\rho^{1-\alpha}F_{n}^{(b)})\\
&=t^{2}(1+\sqrt{d})^{2}(d^{2}-1)-t^{2}\sum_{b=1}^{d+1}
\sum_{n=1}^{d}\mathrm{Tr}(\rho^{\alpha}F_{n}^{(b)}\rho^{1-\alpha}F_{n}^{(b)}),
\end{split}
\end{equation}
where in the last equality, we have used the fact that $\sum_{b=1}^{d+1}\sum_{n=1}^{d}\mathrm{Tr}[(F_{n}^{(b)})^{2}\rho]=(1+\sqrt{d})^{2}(d^{2}-1)$ \cite{ChenT}.

Noting that $\sum_{n=1}^{d-1}F_{n,b}=F^{(b)}$, we have
\begin{equation}\label{t2}
\begin{split}
\sum_{b=1}^{d+1}\sum_{n=1}^{d}\mathrm{Tr}(\rho^{\alpha}F_{n}^{(b)}\rho^{1-\alpha}F_{n}^{(b)})&=
\sum_{b=1}^{d+1}\sum_{n=1}^{d-1}\mathrm{Tr}\{\rho^{\alpha}[F^{(b)}-(d+\sqrt{d})F_{n,b}]
\rho^{1-\alpha}[F^{(b)}-(d+\sqrt{d})F_{n,b}]\}\\
&\quad +(1+\sqrt{d})^{2}\sum_{b=1}^{d+1}\mathrm{Tr}(\rho^{\alpha}F^{(b)}\rho^{1-\alpha}F^{(b)})\\
&=(d+\sqrt{d})^{2}\sum_{b=1}^{d+1}\sum_{n=1}^{d-1}\mathrm{Tr}(\rho^{\alpha}F_{n,b}\rho^{1-\alpha}F_{n,b}).
\end{split}
\end{equation}

Taking into account that $\sum_{b=1}^{d+1}\sum_{n=1}^{d-1}(F_{n,b})^{2}=(d-1/d)I$ \cite{LuoBZ}, we have
\begin{equation}
\begin{split}
\sum_{b=1}^{d+1}\sum_{n=1}^{d-1}I_{\alpha}(\rho,F_{n,b})
&=\sum_{b=1}^{d+1}\sum_{n=1}^{d-1}\{\mathrm{Tr}[(F_{n,b})^{2}\rho]-\mathrm{Tr}(\rho^{\alpha}F_{n,b}\rho^{1-\alpha}F_{n,b})\}\\
&=d-\frac{1}{d}-\sum_{b=1}^{d+1}\sum_{n=1}^{d-1}\mathrm{Tr}(\rho^{\alpha}F_{n,b}\rho^{1-\alpha}F_{n,b}).
\end{split}
\end{equation}
Nevertheless, it has been proved that \cite{Lix}
\begin{equation}
\sum_{b=1}^{d+1}\sum_{n=1}^{d-1}I_{\alpha}(\rho,F_{n,b})
=d-\mathrm{Tr}(\rho^{\alpha})\mathrm{Tr}(\rho^{1-\alpha}).
\end{equation}
Therefore, we obtain
\begin{equation}\label{t3}
\sum_{b=1}^{d+1}\sum_{n=1}^{d-1}\mathrm{Tr}(\rho^{\alpha}F_{n,b}\rho^{1-\alpha}F_{n,b})=\mathrm{Tr}(\rho^{\alpha})\mathrm{Tr}(\rho^{1-\alpha})-\frac{1}{d}.
\end{equation}
Combining Eqs. (\ref{q1}), (\ref{t2}) and (\ref{t3}), we have
\begin{equation}
Q_{\alpha}(\rho,\mathcal{P}_{\mathrm{MUM}})=\frac{\kappa d-1}{d-1}[d-\mathrm{Tr}(\rho^{\alpha})\mathrm{Tr}(\rho^{1-\alpha})].
\end{equation}

Here it is interesting that this quantity $Q_{\alpha}(\rho,\mathcal{P}_{\mathrm{MUM}})$ is tightly related to a measure of quantum uncertainty based on averaging WYD information, which is defined by \cite{Lix}
\begin{equation}
Q_{\alpha}(\rho):=\sum_{i=1}^{d^{2}}I_{\alpha}(\rho,H_{i})=d-\mathrm{Tr}(\rho^{\alpha})\mathrm{Tr}(\rho^{1-\alpha}),
\end{equation}
where $\{H_{i}\}$ is any complete orthogonal set of observables.
One can easily seen that $Q_{\alpha}(\rho,\mathcal{P}_{\mathrm{MUM}})=\frac{\kappa d-1}{d-1}Q_{\alpha}(\rho)$. Moreover, these two quantities are equivalent
when a complete set of MUBs is taken into account, since $\kappa=1$ at this point.

From (\ref{CQ}), we have the following conclusion.

{\bf [Theorem 1]} The average coherence of a state $\rho$ with respect to the $\mathcal{P}_{\mathrm{MUM}}=\{\mathcal{P}^{(b)}\}_{b=1}^{d+1}$ with parameter $\kappa$
is given by
\begin{equation}\label{thm1}
\mathcal{C}(\rho,\mathcal{P}_{\mathrm{MUM}})=\frac{\kappa d-1}{d^{2}-1}[d-(\mathrm{Tr}\sqrt{\rho})^{2}].
\end{equation}

When $\kappa=1$, $\mathcal{P}_{\mathrm{MUM}}$ gives rise to a complete set of MUBs, and in this case $\mathcal{C}(\rho,\mathcal{P}_{\mathrm{MUM}})=\frac{1}{d+1}[d-(\mathrm{Tr}\sqrt{\rho})^{2}]=\mathcal{C}_{\mathrm{MUB}}(\rho)$.
Otherwise, the average coherence with respect to MUMs is always strictly less than the one with respect to MUBs. Consequently, one has the following order relations,
\begin{equation}
\mathcal{C}(\rho,\mathcal{P}_{\mathrm{MUM}})\leq\mathcal{C}_{\mathrm{MUB}}(\rho)=\mathcal{C}_{\mathcal{U}}(\rho)<\mathcal{C}_{\mathrm{max}}(\rho).
\end{equation}
Moreover, it can be seen that
\begin{equation}
\mathcal{C}(\rho,\mathcal{P}_{\mathrm{MUM}})=\frac{d(\kappa d-1)}{d^{2}-1}\mathcal{C}_{\mathrm{max}}(\rho),
\end{equation}
which implies that
\begin{equation}
\frac{\mathcal{C}(\rho,\mathcal{P}_{\mathrm{MUM}})}
{\mathcal{C}_{\mathrm{max}}(\rho)}\rightarrow\kappa
\end{equation}
when $d\rightarrow\infty$.
This means that for high dimensional quantum systems, the ``closeness" of the average coherence with respect to MUMs to the maximal coherence depends heavily on the parameter $\kappa$, and $\mathcal{C}(\rho,\mathcal{P}_{\mathrm{MUM}})$ gets closer to the maximum coherence of $\rho$ when $\kappa$ increases.

\section{Average coherence with respect to general SIC measurements}

In this section, we consider the average coherence of a state with respect to general SIC measurements.
A set of $d^{2}$ positive-semidefinite operators $\{P_{k}\}_{k=1}^{d^{2}}$ on $\mathcal{H}_{d}$ is said to be a general SIC measurements, if \\
\indent(1) $\sum_{k=1}^{d^{2}}P_{k}=\mathbf{1}$,\\
\indent(2) $\mathrm{Tr}(P_{k}^{2})=a,~ \mathrm{Tr}(P_{k}P_{l})=\frac{1-da}{d(d^{2}-1)},~\forall k,l\in\{1,2,\ldots,d^{2}\},~k\neq l$,\\
where $a$ is the efficiency parameter satisfying $\frac{1}{d^{3}}<a\leq\frac{1}{d^{2}}$.
$a={1}/{d^{2}}$ if and only if all $P_{k}$ are rank one projectors, which gives rise to a SIC-POVM.
Like MUBs, the existence of SIC-POVMs in arbitrary dimension $d$ is also an open problem.
It has been only proved that there exist SIC-POVMs for a number of low-dimensional cases (see \cite{SG} and the references therein).
However, there always exist a general SIC measurements for arbitrary $d$, which can be constructed explicitly \cite{GK2}.
Let $\{F_{k}\}_{k=1}^{d^{2}-1}$ be a set of $d^{2}-1$ Hermitian, traceless operators acting on $\mathcal{H}_{d}$,
satisfying $\mathrm{Tr}(F_{k}F_{l})=\delta_{k,l}$. Define $F=\sum_{k=1}^{d^{2}-1}F_{k}$. Then the $d^{2}$ operators \\
\begin{equation}
\begin{split}
P_{k}&=\frac{1}{d^{2}}I+t[F-d(d+1)F_{k}],~~~k=1,2,\ldots,d^{2}-1,\\
P_{d^{2}}&=\frac{1}{d^{2}}I+t(d+1)F,
\end{split}
\end{equation}
form a general SIC measurements.
Here $t$ should be chosen such that $P_{k}\geq0$, and the parameter $a$ is given by
\begin{equation}\label{a}
a=\frac{1}{d^{3}}+t^{2}(d-1)(d+1)^{3}
\end{equation}
from the construction.

We now define the average coherence of a state $\rho$ with respect to a general SIC measurements $\{P_{k}\}_{k=1}^{d^{2}}$ with the parameter $a$ as follows,
\begin{equation}
\mathcal{C}(\rho,\mathcal{P}_{\mathrm{GSM}}):=Q(\rho,\mathcal{P}_{\mathrm{GSM}})
=\sum_{k=1}^{d^{2}}I(\rho,P_{k}).
\end{equation}

Before computing $\mathcal{C}(\rho,\mathcal{P}_{\mathrm{GSM}})$, we first calculate $Q_{\alpha}(\rho,\mathcal{P}_{\mathrm{GSM}}):=\sum_{k=1}^{d^{2}}I_{\alpha}(\rho,P_{k})$.
$\mathcal{C}(\rho,\mathcal{P}_{\mathrm{GSM}})$ can be obtained immediately by setting $\alpha=1/2$.
Note that
\begin{equation}\label{qp}
\begin{split}
Q_{\alpha}(\rho,\mathcal{P}_{\mathrm{GSM}})&=\sum_{k=1}^{d^{2}}I_{\alpha}(\rho,P_{k})\\
&=\sum_{k=1}^{d^{2}}\mathrm{Tr}[(P_{k})^{2}\rho]-\sum_{k=1}^{d^{2}}\mathrm{Tr}(\rho^{\alpha}P_{k}\rho^{1-\alpha}P_{k})\\
&=ad-\sum_{k=1}^{d^{2}}\mathrm{Tr}(\rho^{\alpha}P_{k}\rho^{1-\alpha}P_{k}),
\end{split}
\end{equation}
where we have used the fact that $\sum_{k=1}^{d^{2}}\mathrm{Tr}[(P_{k})^{2}\rho]=ad$ \cite{ChenT}.

On the other hand, taking into account that $\sum_{k=1}^{d^{2}-1}F_{k}=F$, we have
\begin{equation}\label{tp}
\begin{split}
\sum_{k=1}^{d^{2}}\mathrm{Tr}(\rho^{\alpha}P_{k}\rho^{1-\alpha}P_{k})&=
\sum_{k=1}^{d^{2}-1}\mathrm{Tr}\left\{\rho^{\alpha}\left[\frac{1}{d^{2}}I+t(F-d(d+1)F_{k})\right]\rho^{1-\alpha}
\left[\frac{1}{d^{2}}I+t(F-d(d+1)F_{k})\right]\right\}\\
&\quad +\mathrm{Tr}\left\{\rho^{\alpha}\left[\frac{1}{d^{2}}I+t(d+1)F\right]\rho^{1-\alpha}\left[\frac{1}{d^{2}}I+t(d+1)F\right]\right\}\\
&=\frac{1}{d^{2}}+t^{2}d^{2}(d+1)^{2}\sum_{k=1}^{d^{2}-1}\mathrm{Tr}(\rho^{\alpha}F_{k}\rho^{1-\alpha}F_{k})\\
&=\frac{1}{d^{2}}+t^{2}d^{2}(d+1)^{2}\left(\mathrm{Tr}(\rho^{\alpha})
\mathrm{Tr}(\rho^{1-\alpha})-\frac{1}{d}\right),
\end{split}
\end{equation}
where the last equality follows from (\ref{t3}).
Combining Eqs. (\ref{qp}), (\ref{tp}) and the relation between the parameters $t$ and $a$ (\ref{a}), we have
\begin{equation}
Q_{\alpha}(\rho,\mathcal{P}_{\mathrm{GSM}})=\frac{ad^{3}-1}{d(d^{2}-1)}
(d-\mathrm{Tr}(\rho^{\alpha})\mathrm{Tr}(\rho^{1-\alpha})).
\end{equation}
Therefore we obtain the following theorem:

{\bf [Theorem 2]} The average coherence with respect to a general SIC measurements with the parameter $a$ is given by
\begin{equation}
\mathcal{C}(\rho,\mathcal{P}_{\mathrm{GSM}})=\frac{ad^{3}-1}{d(d^{2}-1)}[d-(\mathrm{Tr}\sqrt{\rho})^{2}].
\end{equation}

When $a=1/d^{2}$, $\mathcal{P}_{\mathrm{GSM}}$ reduces to SIC-POVM. Then we have the average coherence of a state $\rho$ with respect to a SIC-POVM,
\begin{equation}
\mathcal{C}_{\mathrm{SIC}}(\rho)=\frac{1}{d(d+1)}[d-(\mathrm{Tr}\sqrt{\rho})^{2}].
\end{equation}

It is interesting to find the relations among $\mathcal{C}_{\mathrm{SIC}}(\rho)$, $\mathcal{C}_{\mathrm{MUB}}(\rho)$ and $\mathcal{C}_{\mathrm{max}}(\rho)$.
Remarkably one sees that $\mathcal{C}_{\mathrm{MUB}}(\rho)=d\mathcal{C}_{\mathrm{SIC}}(\rho)$.
Thus the average coherence of a state provides an operational link between MUBs and SIC-POVMs.
This is also the case between MUMs and general SIC measurements, i.e.,
\begin{equation}
\mathcal{C}(\rho,\mathcal{P}_{\mathrm{MUM}})=\frac{\kappa d^{2}-d}{ad^{3}-1}\mathcal{C}(\rho,\mathcal{P}_{\mathrm{GSM}}),
\end{equation}
where the constant multiple depends on the parameters $\kappa$ and $a$.
Furthermore, it is obvious that $\mathcal{C}_{\mathrm{SIC}}(\rho)=\frac{1}{d+1}\mathcal{C}_{\mathrm{max}}(\rho)$,
which implies that
\begin{equation}
\frac{\mathcal{C}_{\mathrm{SIC}}(\rho)}{\mathcal{C}_{\mathrm{max}}(\rho)}=\frac{1}{d+1}\rightarrow 0
\end{equation}
when $d\rightarrow\infty$.
That is to say, for high dimensional systems, $\mathcal{C}_{\mathrm{SIC}}(\rho)$ is much less than the maximal coherence, which is quite different from the case of $\mathcal{C}_{\mathrm{MUB}}(\rho)$.

As an example, let us consider an arbitrary pure state $\rho$.
Simple calculation shows that $\mathcal{C}_{\mathrm{MUB}}(\rho)=\frac{d-1}{d+1}$, $\mathcal{C}_{\mathrm{max}}(\rho)=\frac{d-1}{d}$,
and $\mathcal{C}_{\mathrm{SIC}}(\rho)=\frac{d-1}{d(d+1)}$.
Hence, one can see that $\mathcal{C}_{\mathrm{MUB}}(\rho)$ is almost the maximal,
while $\mathcal{C}_{\mathrm{SIC}}(\rho)$ approaches to the minimum coherence as $d$ increases, see Fig. 1. In this sense, $\mathcal{C}_{\mathrm{SIC}}(\rho)$ and $\mathcal{C}_{\mathrm{MUB}}(\rho)$ can be viewed as dual quantities to some extent in high dimensional systems.
It is noteworthy that the above discussion is based on the assumption that there exist complete sets of MUBs and SIC-POVMs for arbitrary $d$.
However, these results also apply to $\mathcal{C}(\rho,\mathcal{P}_{\mathrm{GSM}})$,
since $\mathcal{C}(\rho,\mathcal{P}_{\mathrm{GSM}})\leq\mathcal{C}_{\mathrm{SIC}}(\rho)$ due to the range of the parameter $a$.

\begin{figure}
\centering
\includegraphics[width=7cm]{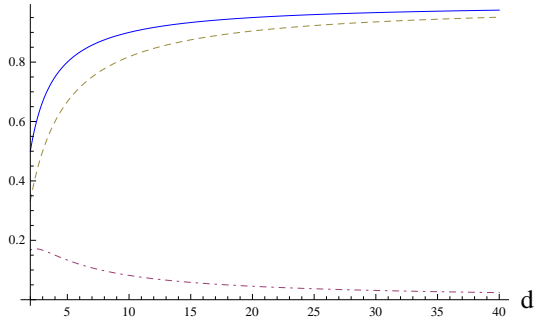}
\caption{The blue solid line is $\mathcal{C}_{\mathrm{max}}(\rho)$,
the dashed line is $\mathcal{C}_{\mathrm{MUB}}(\rho)$, and the dot-dashed line is $\mathcal{C}_{\mathrm{SIC}}(\rho)$.}
\end{figure}

\section{Conclusion}

In summary, we have studied the average coherence with respect to complementary measurements.
By evaluating the average coherence associated with MUMs and general SIC measurements, respectively, we have also established the relations among these quantities and the maximal coherence of quantum states. It has been shown that, for high dimensional systems, the quantity $\mathcal{C}(\rho,\mathcal{P}_{\mathrm{MUM}})$ gets closer to the maximal coherence as the parameter in $\mathcal{P}_{\mathrm{MUM}}$ increases.
However, this is not the case for $\mathcal{P}_{\mathrm{GSM}}$.
Even for a SIC-POVM, the quantity $\mathcal{C}_{\mathrm{SIC}}(\rho)$ approaches to zero when $d$ becomes large. The reasons behind these results are worthy of investigation. One may conjecture that it is related to the number of measurements constituting a POVM.
Our results can offer insight into quantum coherence and complementary measurements.
It would be also interesting to study the measurement-based coherence measure for other types of measurements, and their relations among $\mathcal{C}(\rho,\mathcal{P}_{\mathrm{MUM}})$, $\mathcal{C}(\rho,\mathcal{P}_{\mathrm{GSM}})$ and $\mathcal{C}_{\mathrm{max}}(\rho)$.

\vspace{2.5ex}
\noindent{\bf Acknowledgments}\, \,
This work is supported by the National Natural Science Foundation of China under Grant Nos. 11805143 and 11675113, and Beijing Municipal Commission of Education (KZ201810028042).

\end{document}